\documentclass[12pt]{iopart}

\expandafter\let\csname equation*\endcsname\relax
\expandafter\let\csname endequation*\endcsname\relax
\usepackage{amsmath}
\usepackage{graphicx}
\usepackage{lineno}
\usepackage{diagbox}
\usepackage{hyperref}
\hypersetup{colorlinks,linkcolor={blue},citecolor={blue},urlcolor={blue}}
\usepackage{tcolorbox}
\usepackage{gensymb}
\usepackage{siunitx}
\usepackage{amstext}
\usepackage{multicol}
\usepackage{subcaption}
\usepackage{indentfirst}
\usepackage{booktabs}
\begin{document}


\title[]{Impact of Azimuthal Magnetic Field Inhomogeneity on Hall Thruster high-frequency azimuthal instability via 2D radial-azimuthal PIC simulations}

\author{Zhijun Zhou\textsuperscript{1}, Xin Luo\textsuperscript{2}, Yinjian Zhao\textsuperscript{1,*} and Daren Yu\textsuperscript{1}}
\ead{zhaoyinjian@hit.edu.cn}
\address{\textsuperscript{1}School of Energy Science and Engineering, Harbin Institute of Technology, Harbin 150001, People’s Republic of China}
\address{\textsuperscript{2}College of Aerospace and Civil Engineering, Harbin Engineering University, Harbin 150001, People’s Republic of China}

\vspace{10pt}
\begin{indented}
\item[]\today
\end{indented}

\begin{abstract}
For the SPT-type Hall thrusters, the magnetic structure with magnetic
conductive columns leads inherent to azimuthally inhomogeneous magnetic configurations. 
This azimuthal magnetic inhomogeneity may impact electron 
azimuthal closed-drift motion and cross-field transport characteristics.
This study systematically investigates the effects of azimuthal magnetic field gradient on high-frequency azimuthal instability and
associated anomalous electron transport through 2D radial-azimuthal Particle-in-Cell (PIC) simulations.
The results reveal dual mechanisms of magnetic inhomogeneity on
electron cyclotron drift instability (ECDI) characteristics: (1) The
azimuthal drift velocity distribution becomes modulated by the magnetic field inhomogeneity, with increased average drift velocity enhancing ECDI intensity under stronger inhomogeneity; (2) Simultaneously, the ECDI wavenumber spectrum broadens with elevated magnetic inhomogeneity, reducing discrete ECDI spectral peaks.
Under the dual influence of magnetic field inhomogeneity, when the inhomogeneity level is below 5\%, the ECDI saturation amplitude and 
electron cross-field mobility remains largely unchanged.
However, a notable reduction of 13.4\% in ECDI saturation intensity and a 15.7\% decrease in electron mobility are observed when magnetic
field inhomogeneity reaches 10\%.
\end{abstract}

\section{Introduction}

\begin{figure}[ht]
\centering
    \begin{minipage}[h]{0.35\textwidth}
    \centerline{\includegraphics[scale=0.3]{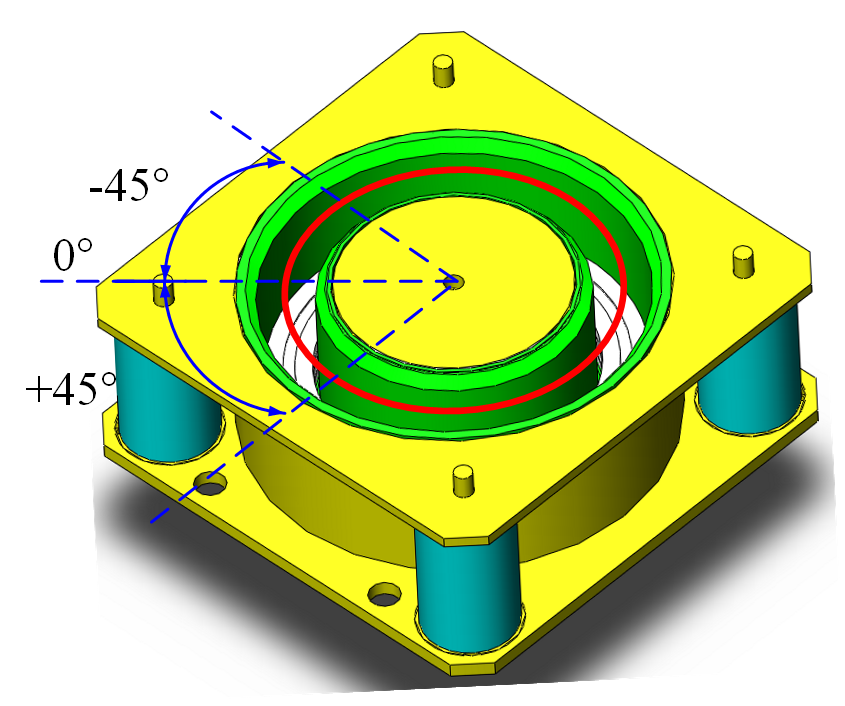}}
    \makebox[1em]{(a)}
    \end{minipage}
    \hspace{35pt}
    \centering
    \begin{minipage}[h]{0.35\textwidth}
    \centerline{\includegraphics[scale=0.5]{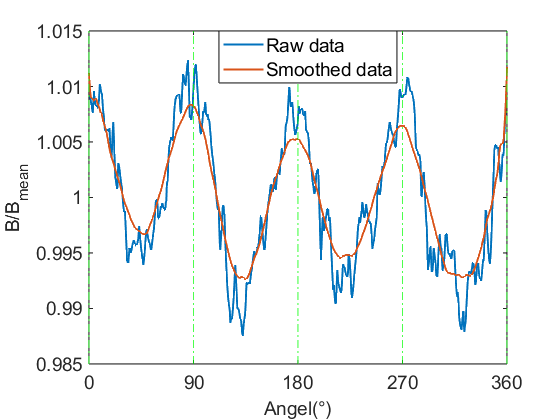}}
    \makebox[1em]{(b)}
    \end{minipage}
\caption{(a) The SPT-type Hall thruster magnetic structure with four magnetic conductive columns, the red line indicates the the median circumference of the channel exit. (b) The radial magnetic field intensity distribution along the red line, obtained by magnetic field simulation software.}
\label{fig:structure}
\end{figure}

The magnetic field configuration of a Hall thruster serves as a critical
physical parameter influencing anomalous electron transport\cite{boeuf2017tutorial}.
In practical applications, four magnetic conductive columns are typically 
employed to replace external coils for enhanced thermal management and reduced thruster weight, as illustrated in the typical SPT-type Hall thruster magnetic circuit structure (Fig.~\ref{fig:structure}\textcolor{blue}{a}).
However, this quadrupole magnetic 
architecture inevitably induces azimuthal inhomogeneity in the radial
magnetic field distribution, as demonstrated in Fig.~\ref{fig:structure}\textcolor{blue}{b}.  
Under the current SPT design paradigm, such azimuthal magnetic inhomogeneity constitutes an inherent phenomenon. 
This inhomogeneous field distribution may disrupt the closed drift motion 
of electrons, impact cross-field electron transport
mechanisms and discharge performance of Hall Thrusters.

Bak et al.\cite{10.1063/5.0067310} conducted experimental and numerical investigations on the influence of magnetic field inhomogeneity on discharge characteristics in a TAL-type Hall thruster.
By cutting the magnetic pole to create extreme azimuthal magnetic field inhomogeneity (the minimum field strength reduced by 35\% compared to the maximum),
their study demonstrated that magnetic field inhomogeneity governs plasma structure formation through its modulation of electron dynamics. 
Specifically, the azimuthal magnetic field gradient enhances the overall axial electron mobility. 
These findings were validated via a 2D axial–azimuthal hybrid simulation, where neutral atoms and ions were treated with PIC methods, while electrons were modeled using fluid methods.
Notably, the observed increase in axial electron mobility reflects alterations in electron transport properties, which are linked to ionization-acceleration dynamics and anomalous electron transport in Hall thrusters. 
While Bak et al.\cite{10.1063/5.0067310} analyzed time-resolved oscillations in axial electron currents using their hybrid model, they did not address the evolution of azimuthal plasma oscillations, which might be due to hybrid model's 
inability of capturing azimuthal instabilities at the scale of electron Larmor gyroradius. 

Current academic consensus identifies \textbf{E}$\times$\textbf{B}-driven azimuthal instabilities as the core mechanism enabling electrons to escape
magnetic confinement, thereby enhancing anomalous electron transport\cite{10.1063/5.0145536}. 
Specifically, the ECDI, characterized by short-wavelength kinetic effects, and modified two-stream instabilities (MTSI) of fluid nature, have been directly linked to enhanced cross-field electron mobility in Hall thrusters.
The frequency of ECDI is approximately 5 MHz, the wavelength is about 1mm, and the direction is along \textbf{E}$\times$\textbf{B}.
ECDI leads to a strong increase in electron-ion friction\cite{Croes_2017}, and evolves into resemble ion acoustic instability because of nonlinear effects\cite{Lafleur_2018,10.1063/1.5017033}. 
The frequency of MTSI is about 1 MHz, the wavelength is several millimeters.
The formation condition of MTSI is related to B and E, and MTSI plays an important role in radial electron heating\cite{10.1063/1.5033896,10.1063/5.0046843}.
The ECDI in Hall thrusters was first identified through 2D axial-azimuthal PIC-MCC simulations by Adam et al\cite{adam2004study}.
Subsequent studies by Ducrocq et al.\cite{Ducrocq2006} derived the plasma
dispersion equation governing ECDI linear growth, while Cavalier et
al.\cite{10.1063/1.4817743} developed numerical iterative algorithms to solve this 
equation, systematically analyzing parametric dependencies on 
azimuthal drift velocity and magnetic field strength.
Lafleur et al.\cite{LafleurI,10.1063/1.4948496} investigated the fundamental physics of the ECDI through 1D reduced PIC simulations and established analytical expressions for linear growth rates using kinetic theory\cite{Lafleur_2017,Lafleur_2018}.
Subsequent studies by Cores et al.\cite{Croes_2017} extended this work to 2D radial-azimuthal PIC simulations, which confirmed the ECDI-induced 
enhancement of the electron-ion friction force.
In parallel, their 2D radial-axial simulations revealed the emergence of the long-wavelength mode.
Notably, Janhunen et al.\cite{10.1063/1.5033896} identified the long-wavelength oscillations in radial-azimuthal PIC simulations as MTSI,
demonstrating its important role in energy inverse-cascade processes and anomalous electron transport during nonlinear evolution phases.

In current research on azimuthal instabilities in Hall thrusters, 2D axial-azimuthal\cite{Charoy_2019} and radial-azimuthal\cite{Villafana_2021} PIC simulation benchmarks have been established using analytical approximations for ionization processes while neglecting ionization collision modeling.
Although the axial-azimuthal benchmark achieves self-consistent axial electric field formation, it overlooks radial wall effects on instability development.
In contrast, the radial-azimuthal model effectively captures multi-mode coupling characteristics of azimuthal instabilities, particularly the interaction between ECDI and MTSI.
Based on 2D benchmarks and the kinetic theory, current studies have elucidated the effects of important parameters such as axial electric field intensity\cite{Reza2024}, radial magnetic field configuration\cite{Reza2023}, and ionic charge\cite{10.1063/5.0122293}  on the azimuthal instability evolution in Hall thrusters.

For SPT-type Hall thrusters, azimuthal magnetic field inhomogeneity represents an inherent phenomenon whose underlying mechanism for enhanced electron mobility remains incompletely understood.
This study aims to reveal the impact of azimuthal magnetic field inhomogeneity on the azimuthal instability evolution and resultant anomalous
electron transport. 
This paper is organized as follows: Sec.~\ref{sec:setup} describes the basic parameter settings of the simulation model and specifies the azimuthal distribution settings of magnetic fields in this study.
Sec.~\ref{sec:simulation result} systematically analyzes the impacts of azimuthal magnetic field inhomogeneity on both plasma parameter distributions and the azimuthal instability spectra.
Sec.~\ref{sec:conclusion} summarizes the principal conclusions and outlines future research directions.

\section{Simulation setup\label{sec:setup}}

\subsection{Basic settings}
\begin{figure}[htb]
	\centering
	\includegraphics[width=0.5\linewidth]{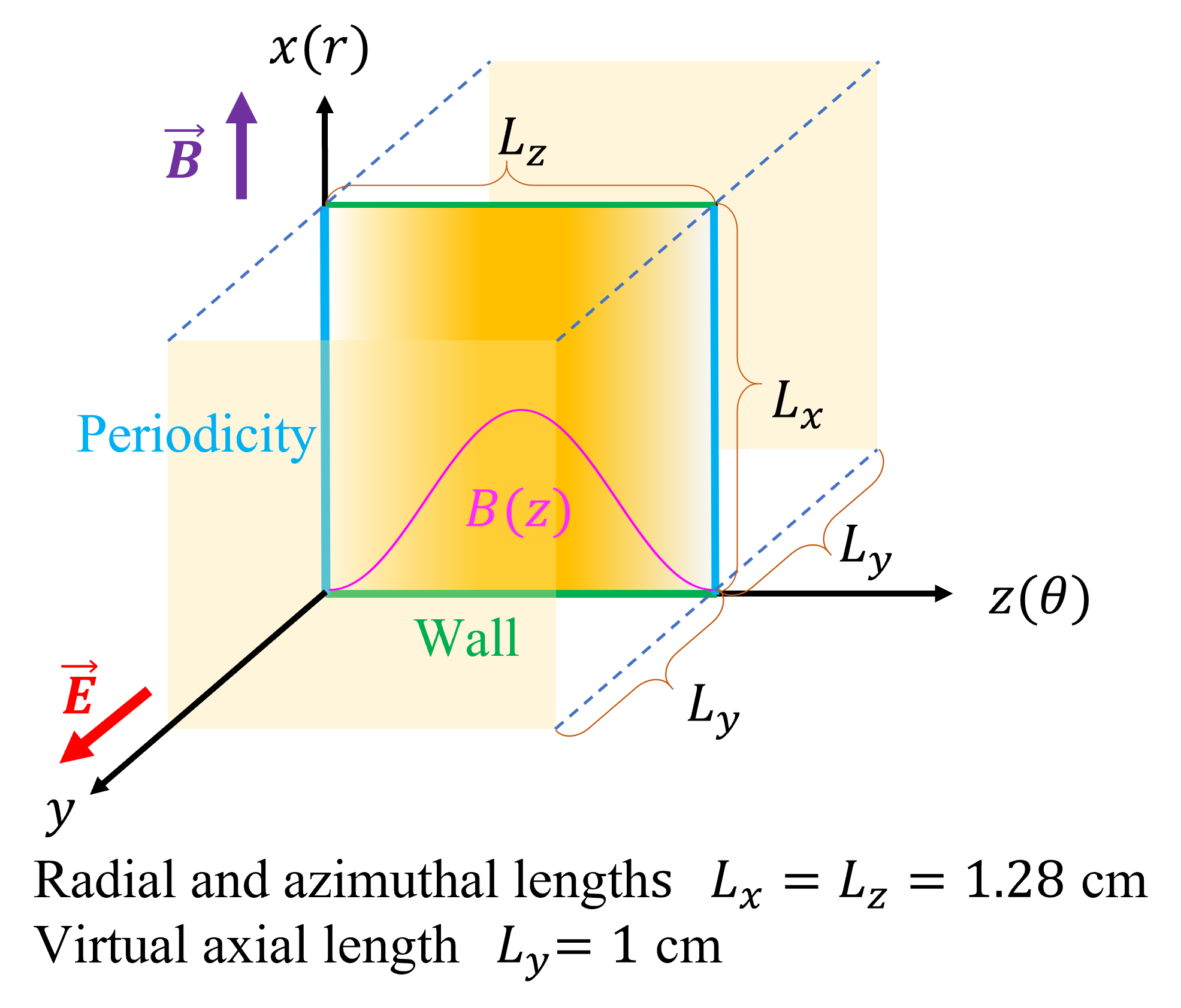}
	\caption{Computational domain setup}
	\label{fig:C_Domain}
\end{figure}
This investigation is based on the 2D radial-azimuthal benchmark established by Villafana et al\cite{Villafana_2021} for modeling azimuthal instability development, with the computational domain illustrated in Fig.~\ref{fig:C_Domain}.
We set the x direction to be the radial direction, the z direction to be the azimuthal direction, and azimuthal curvature is not considered. 
The axial length and azimuthal length are both set to $L_x=L_y$ = 1.28 cm with 256 evenly distributed grids. 
The initial plasma is evenly distributed throughout the simulation domain with density $n_0=5\times10^{16}\mathrm{~m^{-3}}$, the initial electron temperature is $T_e$ = 10 eV with isotropic Maxwellian velocity distribution, the initial ion temperature is $T_i$ = 0.5 eV with isotropic Maxwellian velocity distribution. 
We set the simulation time step  to $\Delta t= 1.5\times10^{-11}$ s and total simulation time to $30\mathrm{~\mu s}$. 
The number of macroparticles per cell is $N_{ppc}$ = 100.
The Dirichlet boundary condition with constant potential $\phi$ = 0 V is used for both the inner and outer walls in the radial direction, and periodic boundary conditions are used in the azimuthal direction.
The radial magnetic induction intensity is set to $\bar{B}$ = 200 Gauss.
Virtual axial boundaries with a thickness of 1 cm is added for injecting and deregistering high-speed particles.
Axial electric field is set to $E_0 = 1.0\times10^4$ V/m.
Simulations were conducted using WarpX, an open-source PIC code developed by Lawrence Berkeley National Laboratory\cite{warpx}. 

As illustrated in Fig.~\ref{fig:structure}\textcolor{blue}{b}, the
magnetic field surrounding each magnetic conductive column exhibits a sinusoidal azimuthal distribution.
To optimize computational efficiency, the simulation domain incorporates a single magnetic conductive column with sinusoidally
distributed magnetic fields spanning ±45° azimuthally, as shown in Fig.~\ref{fig:structure}\textcolor{blue}{a}. 
The implementation of periodic boundary conditions along the azimuthal direction preserves the physical fidelity of the simulation. 
The azimuthal distribution of radial magnetic field is mathematically expressed as
\begin{equation}\label{eq:B_distribution}
B(z)=\bar{B}[1+\alpha_{inh}\sin\left(\frac{2\pi z}{L_{z}}-\frac{\pi}{2}\right)]\quad z\in[0,L_z],
\end{equation}
the azimuthal magnetic field gradient is
\begin{equation}\label{eq:B_gradient}
\nabla B(z)=\frac{2\pi\bar{B}\alpha_{inh}z}{L_{z}}\cos\left(\frac{2\pi z}{L_{z}}-\frac{\pi}{2}\right)\quad   z\in[0,L_z],
\end{equation}
where $\alpha_{inh}=(B_{max}-B_{min})/\bar{B}$
represents the azimuthal inhomogeneity of the radial magnetic field.
The magnetic field gradient in Eq.~\ref{eq:B_gradient} could be modulated by 
adjusting two key parameters: the azimuthal length $L_{z}$, the azimuthal
inhomogeneity of the radial magnetic field $\alpha_{inh}$.
The primary objective of this study is to map the evolution characteristics of azimuthal instabilities across the parametric space defined by the two
key parameters. Therefore, in order to ensure computational fidelity 
against the numerical uncertainty, a parametric independence validation between the simulation results and the azimuthal length needs to be 
verified before advancing to the subsequent stages of the study.

\subsection{Effects of the azimuthal length}

\begin{figure}[ht]
\centering
    \begin{subfigure}[b]{0.35\textwidth}
    \captionsetup{justification=raggedright, singlelinecheck=false}
     \includegraphics[scale=0.28]{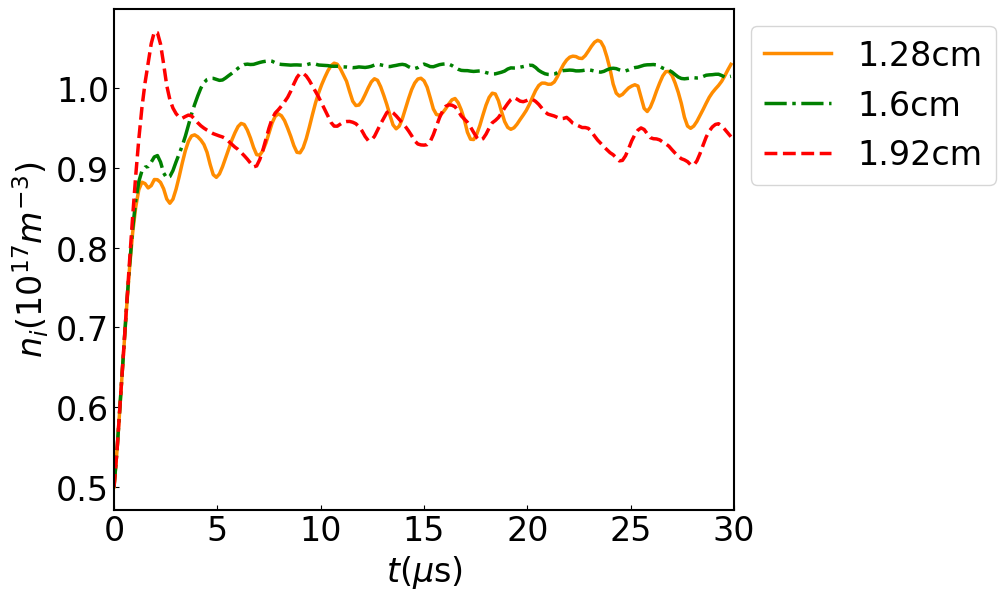}
     \caption{}
    \label{diff_lth_ni}
    \end{subfigure}
    \hspace{38pt}
   \begin{subfigure}[b]{0.35\textwidth}
    \captionsetup{justification=raggedright, singlelinecheck=false}
    \includegraphics[scale=0.28]{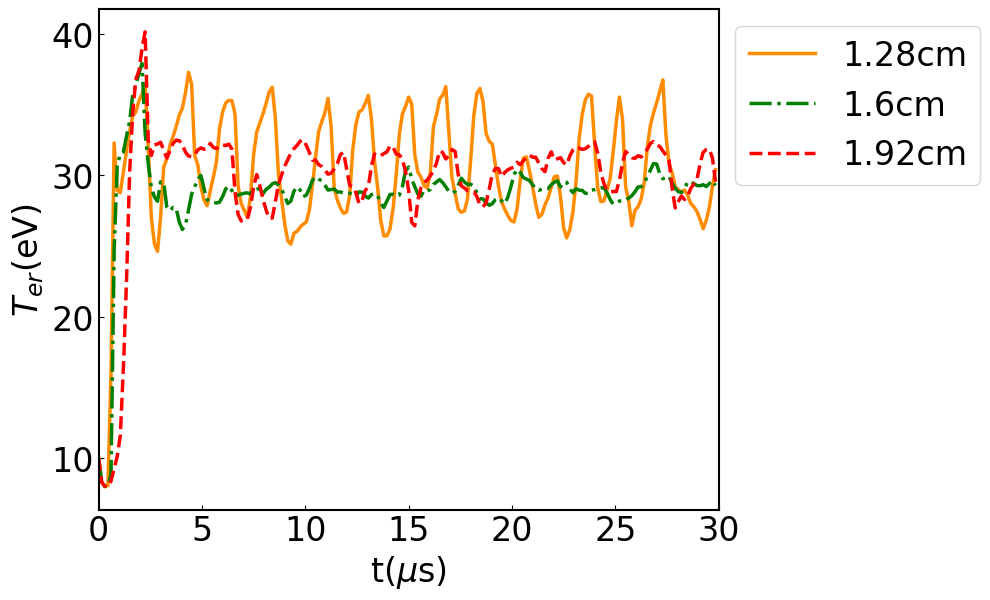}
    \caption{}
    \label{diff_lth_Ter}
    \end{subfigure}
\caption{Comparisions between different azimuthal lengths, (a) ion number density evolving with time, and (b) radial electron temperature evolving with time.}
\label{fig:different_lth}
\end{figure}

Under the condition of maintaining constant azimuthal cell
size, the azimuthal domain length was extended by increasing the 
number of grids.
Three distinct azimuthal lengths (1.28 cm, 1.6 cm, and 1.92 cm) were investigated to examine cross-field discharge characteristics in Hall thrusters. 
Fig.~\ref{fig:different_lth} presents the temporal evolution of ion density 
and radial electron
temperature profiles. The 1.28 cm case exhibits pronounced 
oscillatory behavior in both ion density and electron temperature 
distributions at steady-state, whereas the 1.6 cm case demonstrates relatively stable
temporal profiles.

\begin{figure}[htb]
	\centering
	\includegraphics[width=0.45\linewidth]{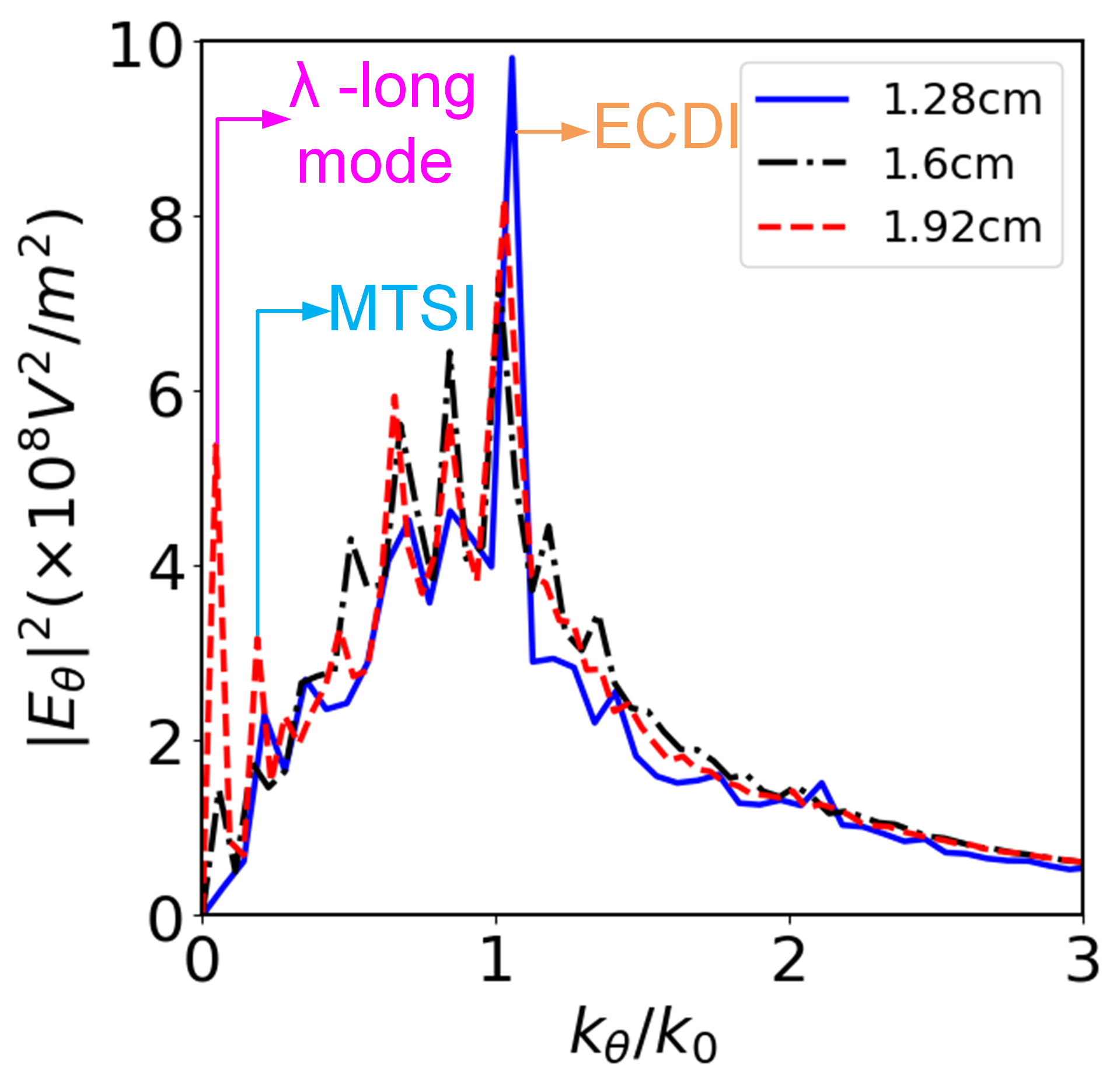}
	\caption{Comparison of time-averaged azimuthal electric field amplitudes at different normalized wavenumbers during 21-27 $\mu$s, $k_0$ = 7035 $\mathrm{m^{-1}}$. And this figure delineates the wavenumbers corresponding to the ECDI, MTSI and $\lambda$-long mode. The three distinct peaks observed between ECDI and MTSI arise from the resonant interaction of multiple instability modes at beat frequencies.}
	\label{fig:EET1D_difflth}
\end{figure}

Fig.~\ref{fig:EET1D_difflth} compares the time-averaged spectral amplitudes of azimuthal electric field fluctuations with varying wavenumbers ($k_ \theta$) during 21-27 $\mu$s. 
For the 1.28 cm case, the ECDI exhibits significantly higher amplitude compared to longer azimuthal lengths. 
Notably, the MTSI intensity increases with the azimuthal length, where the amplitudes at beat frequencies (the term ``beat frequency" comes from the wave-wave interaction theory in the academic monograph of F.F. Chen\cite{chen1984introduction}, where the mutual coupling between harmonic components leads to amplification of the beat wave at the beat frequency) for 1.6 cm and 1.92 cm cases 
surpass those of 1.28 cm, indicating enhanced energy transfer from
ECDI to long-wavelength modes under extended azimuthal dimensions.
As azimuthal length increases, smaller wavenumber components of
azimuthal electric field fluctuations become resolvable.
Specifically, for azimuthal length 1.92 cm,
a prominent $\lambda$-long mode emerges at normalized wavenumber 
$k_{\theta}/k_{0}$ = 0.0435.
This parametric study seems to reveal azimuthal length dependencies:
the shorter azimuthal length preferentially excite high-wavenumber
instabilities, while extended domains
enable physical development of long wavelength modes.
Ref.~\cite{10.1063/1.5139035} also found that the development of long wavelength modes at real Hall thruster channel sizes enhances anomalous
electron transport compared to smaller simulation domain sizes.

The present findings demonstrate that variations in azimuthal length alter the simulated plasma dynamics and the azimuthal instability 
characteristics. 
This pronounced domain-size dependence precludes direct comparisons between simulation results of different azimuthal lengths.
Given the demonstrated capability of the 1.92 cm azimuthal length in 
capturing a broader spectrum of the instability phenomena, the extended azimuthal
length of 1.92 cm was selected as the baseline configuration for subsequent
investigations. 

\subsection{Azimuthal magnetic field inhomogeneity setting}

\begin{figure}[htb]
	\centering
	\includegraphics[width=0.45\linewidth]{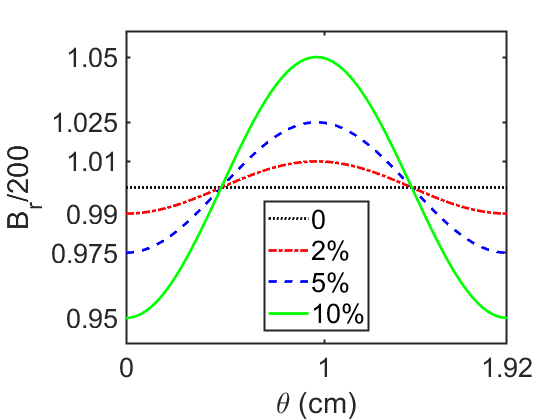}
	\caption{Azimuthal magnetic field configuration of the four simulation cases.}
	\label{fig:azimuthal_cichang}
\end{figure}

As shown in Fig.~\ref{fig:structure}\textcolor{blue}{b}, the magnetic circuit with four magnetic conductive columns of the SPT-100 Hall thruster
exhibits a radial magnetic field with  $\alpha_{inh} \approx$ 2.5\% azimuthal
inhomogeneity under the cold-state condition.  To account for magnetic permeability
variations in magnetic circuit materials under steady-state thermal loading
and Hall current-induced magnetic field perturbations in high-power
discharge regimes, we appropriately augmented the magnetic inhomogeneity levels.
Thus four distinct $\alpha_{inh}$ configurations were implemented: 0\% (ideal homogeneous), 2\%, 5\%, and 10\%, with 
corresponding azimuthal magnetic field distributions illustrated in
Fig.~\ref{fig:azimuthal_cichang}, in order to study the effect of azimuthally inhomogeneous magnetic field on plasma distributions, the 
coupling and evolution of azimuthal instabilities, and the variation
characteristics of anomalous cross-field electron transport.
This study is supposed to provide important insights into cross-field plasma 
dynamics and offer practical guidance for optimizing the magnetic circuit
design in Hall thrusters.

\section{Result analyses\label{sec:simulation result}}
\subsection{Effects on plasma azimuthal profiles\label{sec:plasma dis}}

\begin{figure}[ht]
\centering
    \begin{subfigure}[b]{0.35\textwidth}
    \captionsetup{justification=raggedright, singlelinecheck=false}
     \includegraphics[scale=0.34]{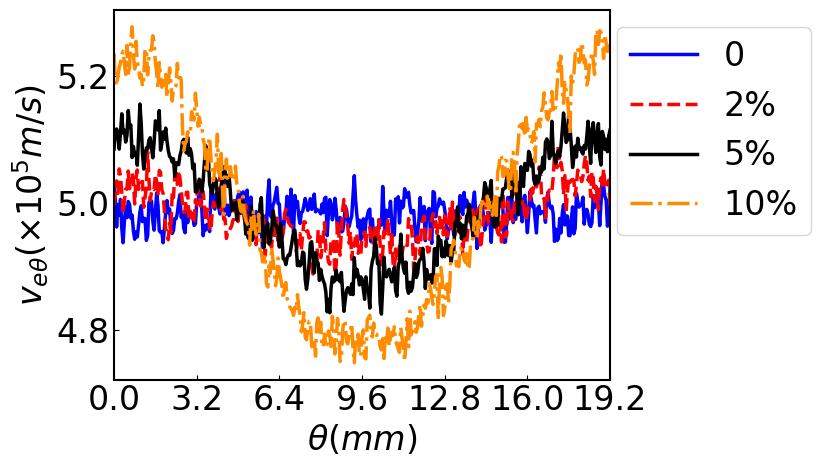}
     \caption{}
    \label{the_the}
    \end{subfigure}
    \hspace{38pt}
   \begin{subfigure}[b]{0.34\textwidth}
    \captionsetup{justification=raggedright, singlelinecheck=false}
    \includegraphics[scale=0.35]{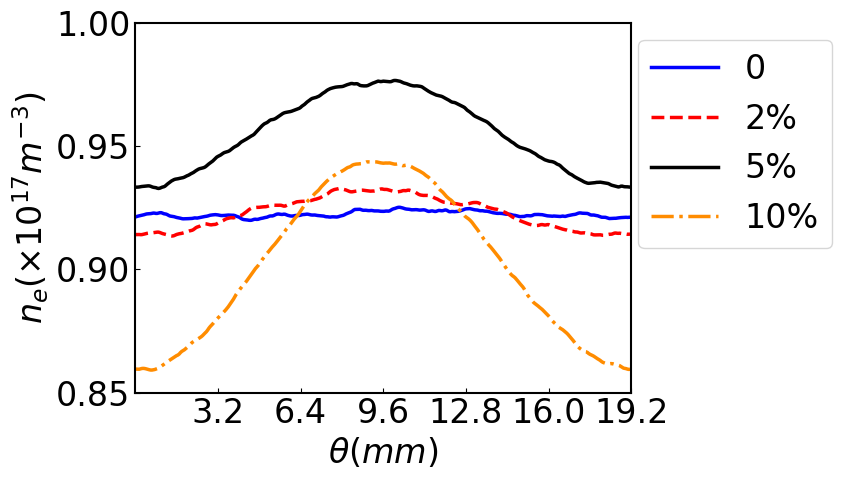}
    \caption{}
    \label{ne}
    \end{subfigure}
\caption{Time-averaged and radially averaged (a) azimuthal electron drift velocity profiles and (b) azimuthal electron density profiles within the 21-27 $\mu$s.}
\label{fig:v_n_azimuthal}
\end{figure}

As delineated in Fig.~\ref{fig:v_n_azimuthal}, the electron azimuthal velocity demonstrates an inverse proportionality to magnetic field intensity.
In regions with high-B, the reduced azimuthal drift velocity leads to
electron accumulation and enhanced confinement capability, ultimately 
forming an azimuthal electron density distribution that mirrors the magnetic field's spatial azimuthal configuration.

\begin{figure}[ht]
\centering
    \begin{subfigure}[b]{0.35\textwidth}
    \captionsetup{justification=raggedright, singlelinecheck=false}
     \includegraphics[scale=0.34]{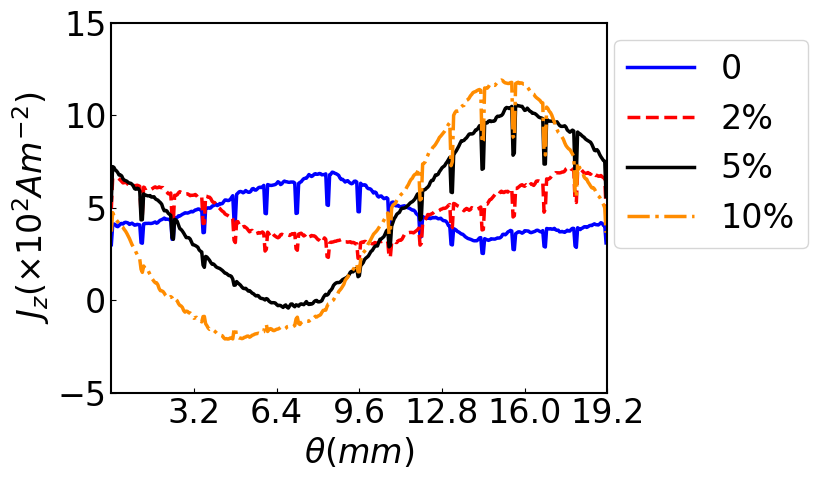}
     \caption{}
    \label{Jz}
    \end{subfigure}
    \hspace{38pt}
   \begin{subfigure}[b]{0.34\textwidth}
    \captionsetup{justification=raggedright, singlelinecheck=false}
    \includegraphics[scale=0.35]{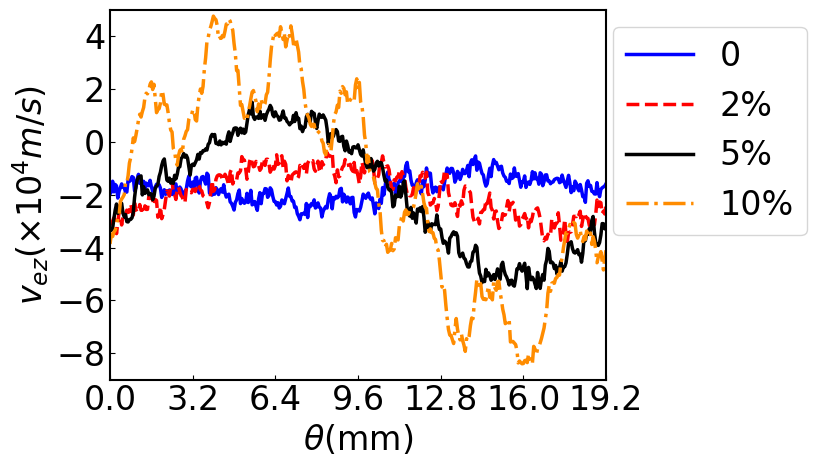}
    \caption{}
    \label{vthe}
    \end{subfigure}
\caption{Time-averaged and radially averaged (a) axial current density profiles and (b) axial electron velocity profiles along azimuthal direction within the 21-27 $\mu$s.}
\label{fig:J_z}
\end{figure}

The azimuthal current density profiles in Fig.~\ref{Jz} exhibit phase 
discrepancy with the electron density distributions in Fig.~\ref{ne}, 
attributable to the coupled effects of the azimuthal magnetic field 
gradient ($\nabla B$) and its induced plasma density gradient ($\nabla n$).
These gradients drive the
azimuthal redistribution of axial electron drift velocity, as quantified in Fig.~\ref{vthe}.
Notably, electron backflow (the direction of the electron axial velocity is the same as the electric field) emerges in the axial positive gradient region (0-9.6 mm) under 5\% and 10\% magnetic field 
gradient, demonstrating the important role of the azimuthal magnetic field gradient in 
governing cross-field electron transport dynamics under these conditions.
The impacts of gradient drifts will be comprehensively discussed in the subsequent section. 

\begin{figure}[htb]
	\centering
	\includegraphics[width=0.4\linewidth]{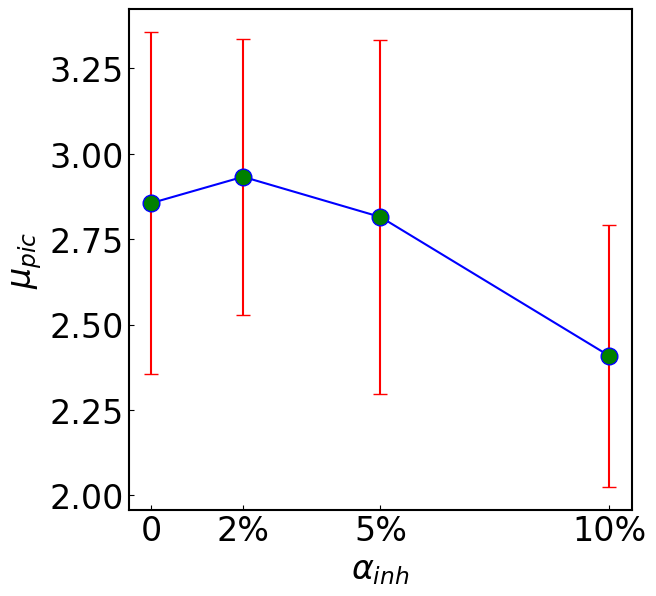}
	\caption{The temporally and spatially averaged axial electron mobility with associated error bars during the 21-27 $\mu$s interval.}
	\label{fig:mu}
\end{figure}

The preceding analysis demonstrates that the magnetic field gradient  modulates axial cross-field electron transport.
To quantify the impact of magnetic field inhomogeneity on electron cross-field transport, we employ axial electron mobility 
as the diagnostic metric, defined as\cite{Tavant_2018}:
\begin{equation}\label{eq:mu}
\mu_{PIC} = -\frac{\langle v_{ez} \rangle}{E_z},
\end{equation}    
where$\langle v_{ez} \rangle$ represents the average axial electron velocity, $E_z$ represents the axial electric field.

Fig.~\ref{fig:mu} delineates the temporally and spatially averaged axial electron mobility with associated error bars during the 21-27 $\mu$s interval. A statistically invariant $\mu_{PIC}$
(±2.5\% variation) is observed for the magnetic field inhomogeneity level $\alpha_{inh} \leq$ 5\%, whereas a pronounced reduction (-15.7\%) emerges at 10\%. 
Changes in electron mobility imply that the magnetic field inhomogeneity may have effects on characterizations of the azimuthal instability.

\subsection{Effects of gradient drifts}\label{gradient_drift}

\begin{figure}[ht]
\centering
    \begin{subfigure}[b]{0.35\textwidth}
    \captionsetup{justification=raggedright, singlelinecheck=false}
     \includegraphics[scale=0.34]{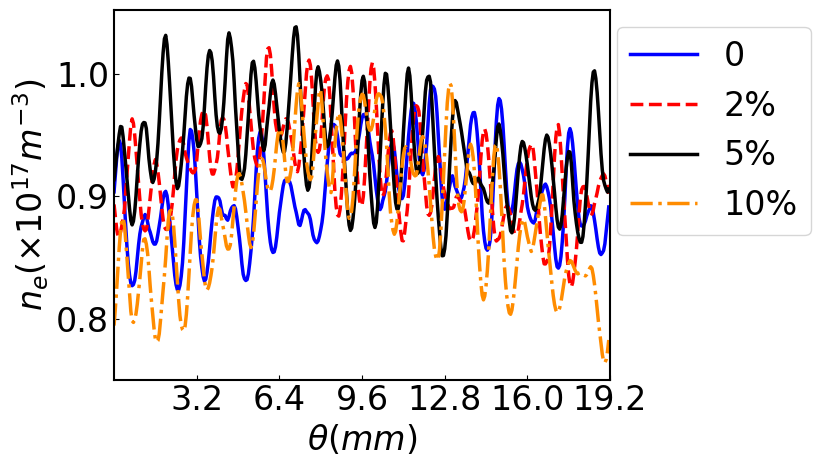}
     \caption{}
    \label{fig:ne24}
    \end{subfigure}
    \hspace{38pt}
   \begin{subfigure}[b]{0.34\textwidth}
    \captionsetup{justification=raggedright, singlelinecheck=false}
    \includegraphics[scale=0.35]{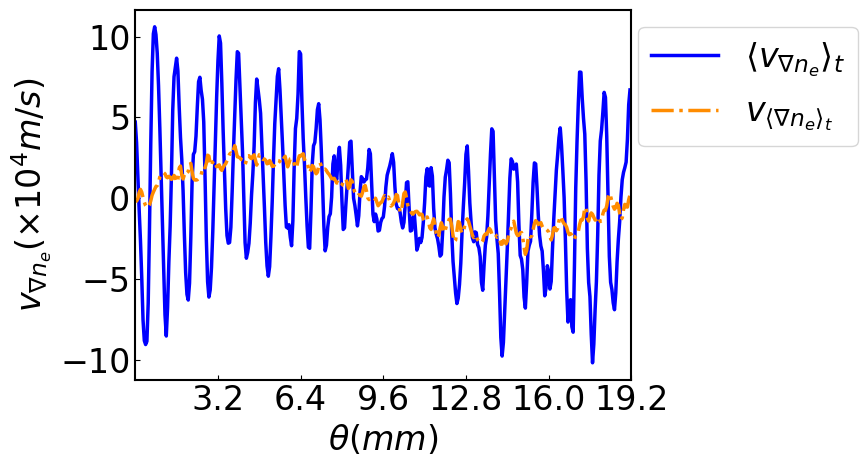}
    \caption{}
    \label{fig:n_drift}
    \end{subfigure}
\caption{(a) Azimuthal electron density profiles at 24 $\mu$s under radial averaging. 
(b) The electron density gradient drift velocity of the case $\alpha_{inh}$ = 10\% obtained with two different methods.}
\label{fig:different_v}
\end{figure}

While the preceding section revealed that the azimuthal magnetic field gradient induces the electron density distribution correlated with the 
magnetic topology, Fig.~\ref{fig:ne24} demonstrates that at discrete 
instants (e.g., 24 $\mu$s shown), high-frequency electrostatic instabilities 
generate small-scale turbulent fluctuations in the electron density, which 
mask the characteristic azimuthal wavelength features of electron density modulation. 
According to single-particle motion theory\cite{frias2016plasma,chen1984introduction}, the electron plasma density gradient drift velocity is expressed as
\begin{equation}\label{eq:v_dn}
\boldsymbol{v}_{\nabla n_e}
= \frac{T_e}{B} \frac{1}{L_n}\boldsymbol{e_y},
\end{equation}   
where $L_n = ( \partial ln(n_e)/ \partial \theta)^{-1}$, $T_e$ (eV) indicates the temperature of the electron, $\boldsymbol{e_y}$ represents a unit vector that is in the same direction as the axial electric field.
Fig.~\ref{fig:n_drift} illustrates the electron density gradient drift velocity obtained by two different methods, where
\begin{equation}\label{eq:meanvdn}
{\langle v_{\nabla n_e} \rangle}_t = \frac{1}{B} \frac{(\frac{T_e}{L_n})_{t_1} + (\frac{T_e}{L_n})_{t_2} + \cdots + (\frac{T_e}{L_n})_{t_n}}{n},
\end{equation}    
\begin{equation}\label{eq:vdn}
v_{{\langle \nabla n_e\rangle}_t}= \frac{1}{B} \frac{(\frac{1}{L_n})_{t_1} + (\frac{1}{L_n})_{t_2} + \cdots + (\frac{1}{L_n})_{t_n}}{n} \frac{(T_e)_{t_1} +(T_e)_{t_2}+ \cdots + (T_e)_{t_n}}{n}.
\end{equation}   
From the comparison in Fig.~\ref{fig:n_drift}, the statistical method in Eq.~\ref{eq:vdn} suppresses the effect of small-wavelength perturbations on 
the electron density gradient. 
The magnetic field gradient drift
velocity is expressed as
\begin{equation}\label{eq:v_dB}
\boldsymbol{v}_{\nabla B}
= \frac{T_e}{B} \frac{1}{L_B}\boldsymbol{e_y},
\end{equation}   
where $L_B = ( \partial ln(B)/ \partial \theta)^{-1}$.

\begin{figure}[ht]
\centering
    \begin{subfigure}[b]{0.35\textwidth}
    \captionsetup{justification=raggedright, singlelinecheck=false}
     \includegraphics[scale=0.34]{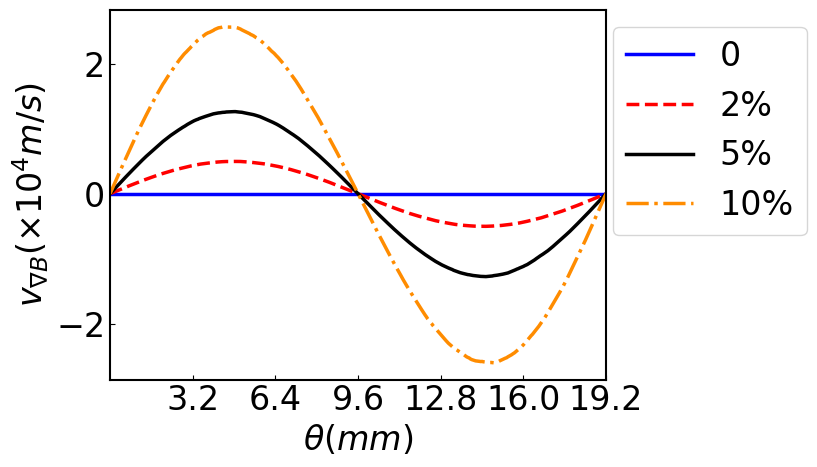}
     \caption{}
    \label{vdB}
    \end{subfigure}
    \hspace{38pt}
   \begin{subfigure}[b]{0.35\textwidth}
    \captionsetup{justification=raggedright, singlelinecheck=false}
    \includegraphics[scale=0.35]{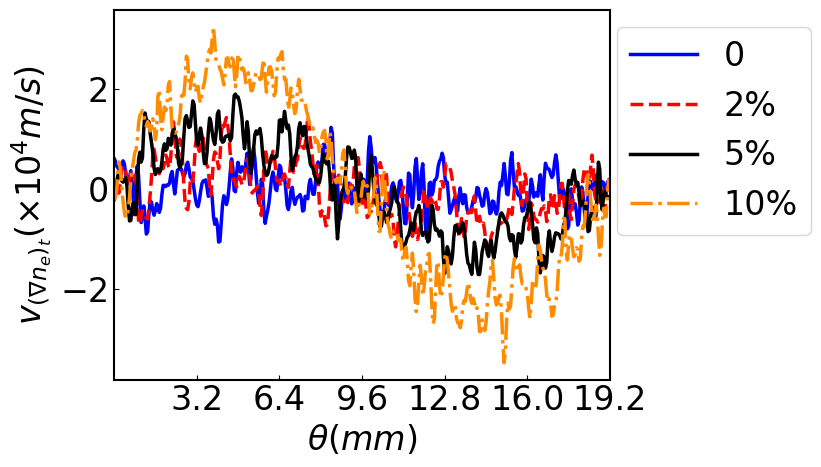}
    \caption{}
    \label{vdn}
    \end{subfigure}
\caption{ (a) The magnetic field gradient drift velocity profiles along azimuthal direction within 21-27 $\mu$s. (b) The electron density gradient drift velocity profiles along azimuthal direction within 21-27 $\mu$s obtained by Eq.~\ref{eq:vdn}.}
\label{fig:vdB}
\end{figure}

\begin{table}[]
	\centering
        \caption{Comparisons between the theoretically calculated gradient drift velocity with the simulated axial electron velocity, $(v_{ez} )_{mean}$ represents the average electron axial velocity of simulations, $v_{\nabla B}$ represents the theoretical magnetic gradient drift velocity, $v_{\nabla n_e}$ represents the theoretical density gradient drift velocity. When calculating the theoretical axial
gradient drift velocity based on the single-particle motion
theory, the electron temperature ($T_e$) and electron density ($n_e$) in the analytical formulation are directly extracted from simulation results.}
	\begin{tabular}{ccccccc}
		\toprule
		velocity($10^{4}\mathrm{m/s}$)      & 0    & 2\%  & 5\% & 10\%  \\ \midrule
Minimum $v_{\nabla B}$  & 0 & -0.5  & -1.27  & -2.6    \\
Maximum $v_{\nabla B}$ & 0 & 0.5 & 1.26 & 2.56 \\
Mean $v_{\nabla B}$ ($\mathrm{m/s}$) & 0 & -12.12 & -24.72 & -97.54 \\
Minimum $v_{\nabla n_e}$ & -1.07  & -1.26   & -1.72 & -3.48 \\
Maximum $v_{\nabla n_e}$ & 1.2& 1.44  & 1.9 & 3.24  \\
Mean $v_{\nabla n_e}$($\mathrm{m/s}$) & 1.83& -6.63  & -19.36 & -93.23  \\
Minimum $v_{ez} - (v_{ez})_{mean}$ & -1.19& -1.85  & -3.46 & -6.37  \\
Maximum $v_{ez} - (v_{ez})_{mean}$ & 1.22& 1.82  & 3.6 & 6.81  \\
\bottomrule
	\end{tabular}
	\label{tab:v_comparison}
\end{table}

The comparison between Fig.~\ref{fig:vdB} and Fig.~\ref{vthe} reveals 
the significant influence of gradient drifts on the azimuthal distribution of the 
axial electron velocity. 
Tab.~\ref{tab:v_comparison} quantifies these effects by contrasting theoretical gradient drift velocities with simulated velocity spatial fluctuation peaks. 
While the statistical method in Eq.~\ref{eq:vdn} attenuates 
the small-wavelength instability-induced density fluctuation, residual
contributions persist, resulting in non-zero density gradient drift
velocities even under the homogeneous magnetic field.
As quantified in Tab.~\ref{tab:v_comparison}, the magnetic gradient 
drift velocity contribution escalates with increasing inhomogeneity: 
29.7\% at $\alpha_{inh}$ = 2\%, 35.9\% at 5\%, and 39.2\% at 10\%. 
The magnetic field gradient further induces azimuthal density
gradients that couple large-scale modulations (characterized by the
magnetic field gradient length $L_B$) with small-scale fluctuations from
azimuthal instabilities.
Therefore, as the magnetic field gradient increases, the density drift
velocity also increases.
From the perspective of the mean velocities of magnetic gradient drift and density gradient drift, 
although increasing inhomogeneity enhances the mean drift velocities opposing the electric field $E_0 \boldsymbol{e_y}$,
the maximum incremental velocity does not exceed 100 m/s, 
thereby exerting negligible effects on the variation in electron mobility.

\begin{figure}[htb]
	\centering
	\includegraphics[width=0.45\linewidth]{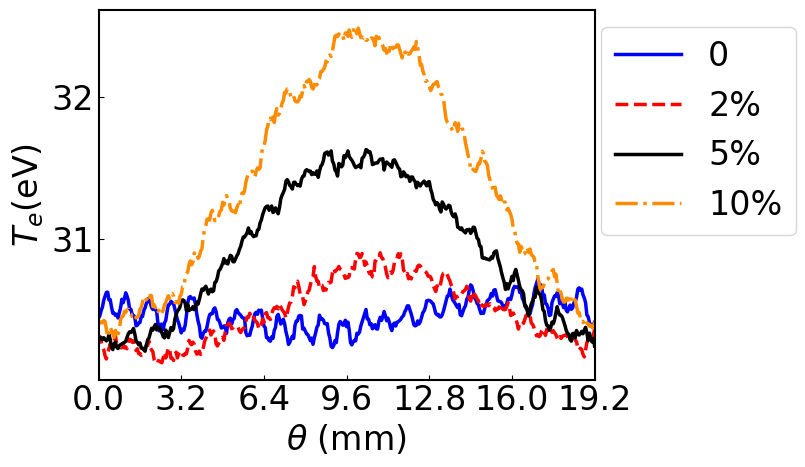}
	\caption{Time-averaged and radially averaged electron temperature profiles along azimuthal direction within 21-27 $\mu$s.}
	\label{fig:Te}
\end{figure}

The drift velocity expressions explicitly incorporate the electron
temperature $T_e$ dependence, with Fig.~\ref{fig:Te} showing the 
azimuthal distributions of time-averaged $T_e$. 
The profile is governed by two mechanisms: (1) Enhanced magnetic
density confines higher-energy electrons, resulting in the $T_e$ profile
spatially correlated with the magnetic field topology. 
(2)The azimuthal electron temperature distribution exhibits asymmetry.
In the negative gradient region (9.6-19.2 mm azimuthal domain), the electron temperature exceeds that in the positive gradient region (0-9.6 mm azimuthal domain). 
This phenomenon originates from gradient drift effects, that
gradient-induced effective forces ($\boldsymbol{F}_{\nabla B} = \mathrm{ k_B T_{e,K}\nabla ln(B) }\boldsymbol{e_y}$ and $\boldsymbol{F}_{\nabla n_e} =\mathrm {k_B T_{e,K}\nabla ln(n_e)} \boldsymbol{e_y}$) modulate axial electron dynamics. 
In the positive gradient region, effective
forces oppose to the axial electric field force -e$\mathrm{E_0} \boldsymbol{e_y}$, enhancing energy 
dissipation. 
Conversely, in the negative gradient region, effective forces aligns with -e$\mathrm{E_0} \boldsymbol{e_y}$, reducing energy 
deposition.
It is evident that the azimuthal magnetic field configuration influences electron energy transport dynamics
within Hall thrusters.
\subsection{Effects on the azimuthal instability}

\begin{figure}[htb]
	\centering
	\includegraphics[width=0.6\linewidth]{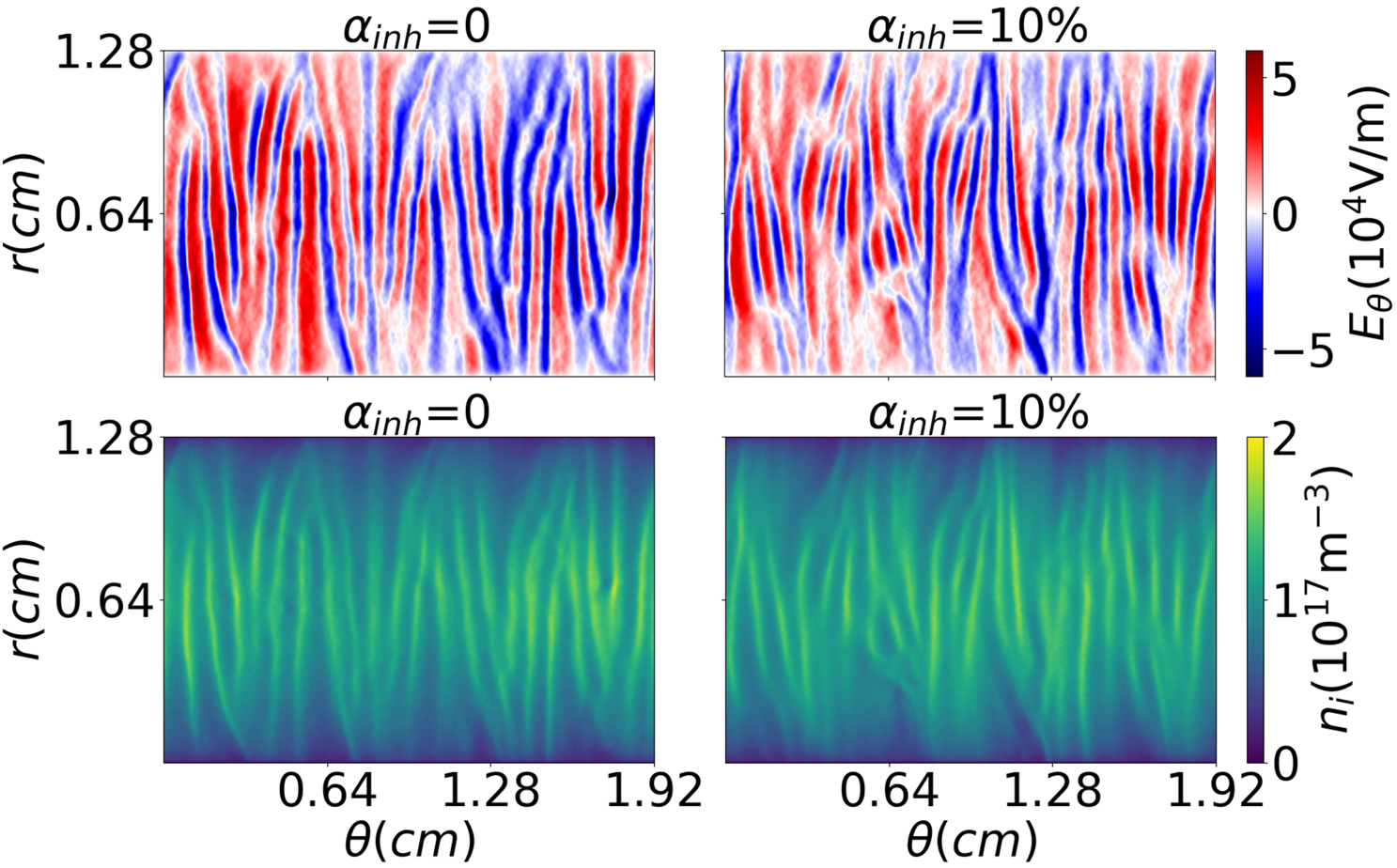}
	\caption{2D maps of the azimuthl electric field and the ion density at 30 $\mu$s.}
	\label{fig:2D}
\end{figure}

Building upon the prior analysis of azimuthal magnetic field gradient
effects on electron dynamics, this section investigates the 
instability modulation mechanisms through spectral characterization of
azimuthal electric field fluctuations.
As illustrated in Fig.~\ref{fig:2D}, the 
azimuthal electric field and ion density fluctuations under varying
magnetic field gradients reveals coupling characteristics between
multiple instabilities.
By the method in Ref.~\cite{Lafleur_2018}, the ECDI wavelength and frequency under
different gradients are estimated to be 0.85-0.95 mm and 4.8-6 MHz.

\subsubsection{Effects on the linear growth stage}
\begin{figure}[htb]
	\centering
	\includegraphics[width=0.99\linewidth]{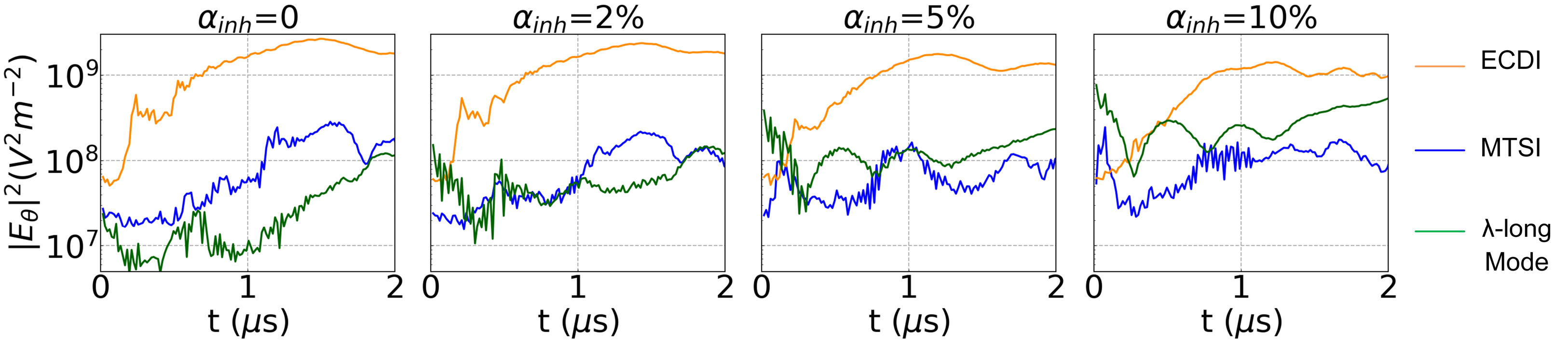}
	\caption{The amplitudes of instability modes within 0-2 $\mu$s.}
	\label{fig:liner}
\end{figure}

\                                                                      
\newline \indent Fig.~\ref{fig:liner} demonstrates that during the initial 0-0.5 $\mu$s, the initial amplitudes of two long-wavelength instability modes (MTSI
and $\lambda$-long mode) increase significantly with increasing  magnetic field gradient. 
Throughout the linear growth phase,
the linear characteristics of ECDI and MTSI gradually diminish due to energy dissipation through inverse energy cascade.
At 2 $\mu$s following the end of linear growth stage, as the magnetic field inhomogeneity intensifies, the ECDI amplitude decreases while 
the $\lambda$-long mode amplitude increases, with minimal variation
observed in MTSI amplitude.

\begin{figure}[htb]
	\centering
	\includegraphics[width=0.45\linewidth]{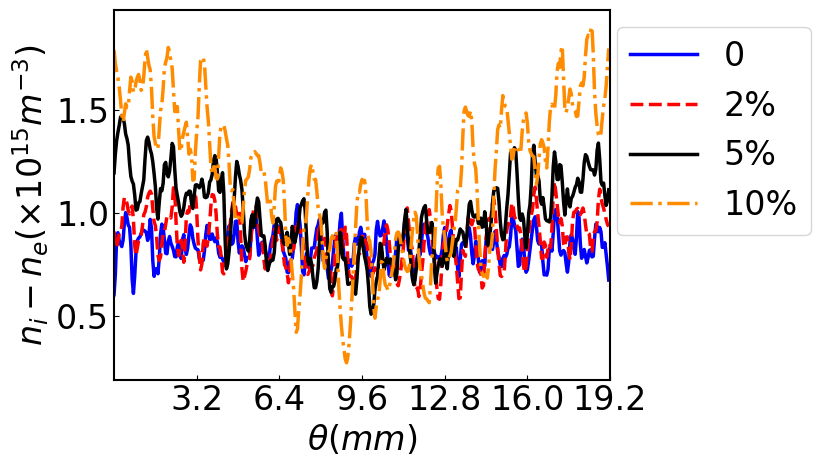}
	\caption{Time-averaged and radially averaged ion-electron density difference profiles along azimuthal direction within 0-0.45 $\mu$s.}
	\label{fig:dn}
\end{figure}

\begin{figure}[htb]
	\centering
	\includegraphics[width=0.6\linewidth]{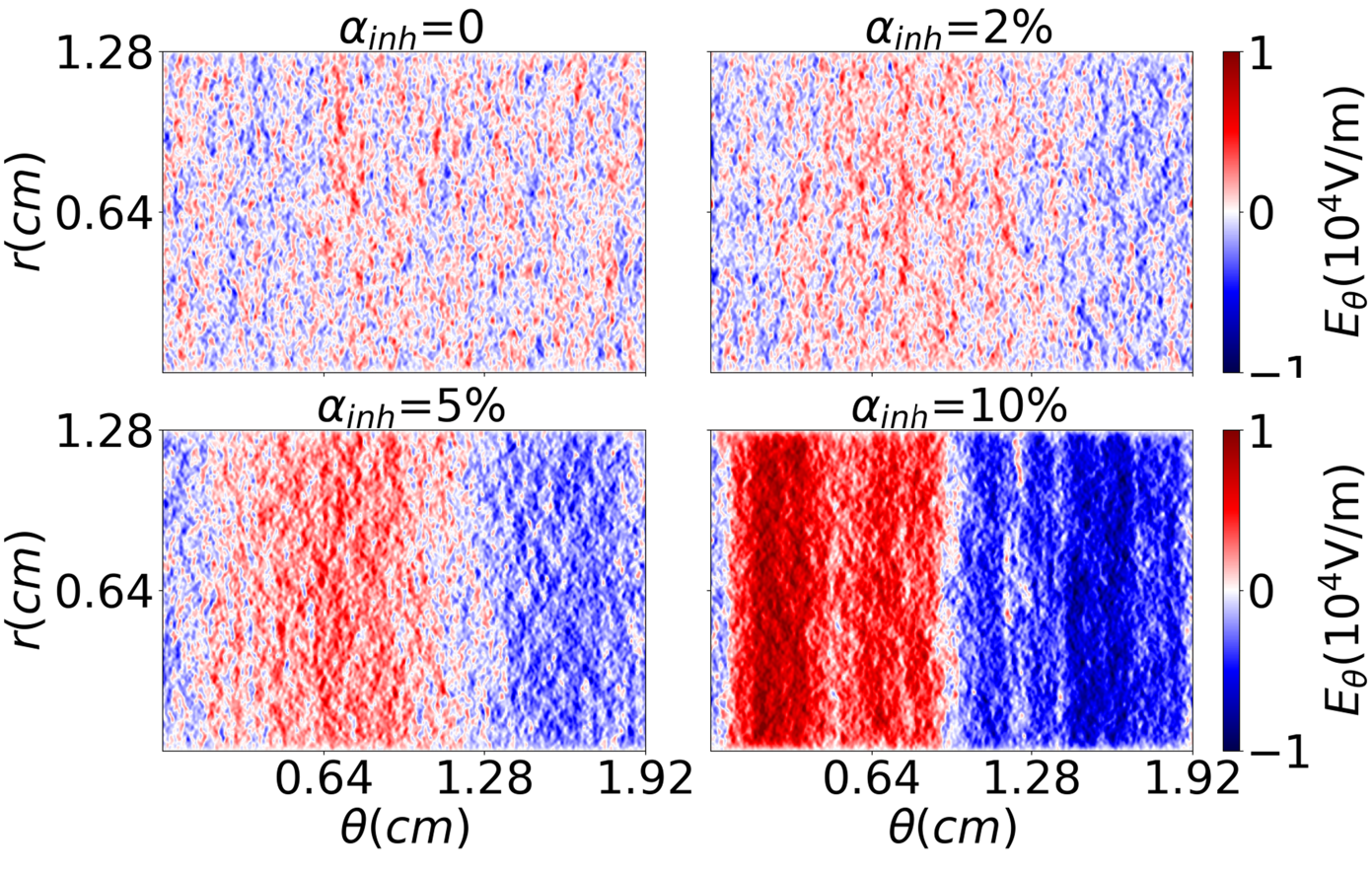}
	\caption{2D maps of the azimuthl electric field at 0.06 $\mu$s.}
	\label{fig:2dEt}
\end{figure}

In the initial simulation phase, electrons rapidly accumulate in high-B regions, while ion density
distributions remain largely unperturbed.
This results in large-scale space charge separation
(shown in Fig.~\ref{fig:dn}) and generates a dominant azimuthal electric field $E_\theta$, whose
intensity exhibits a positive correlation with magnetic field
gradient, as illustrated in Fig.~\ref{fig:2dEt}. 

\begin{figure}[htb]
	\centering
	\includegraphics[width=0.45\linewidth]{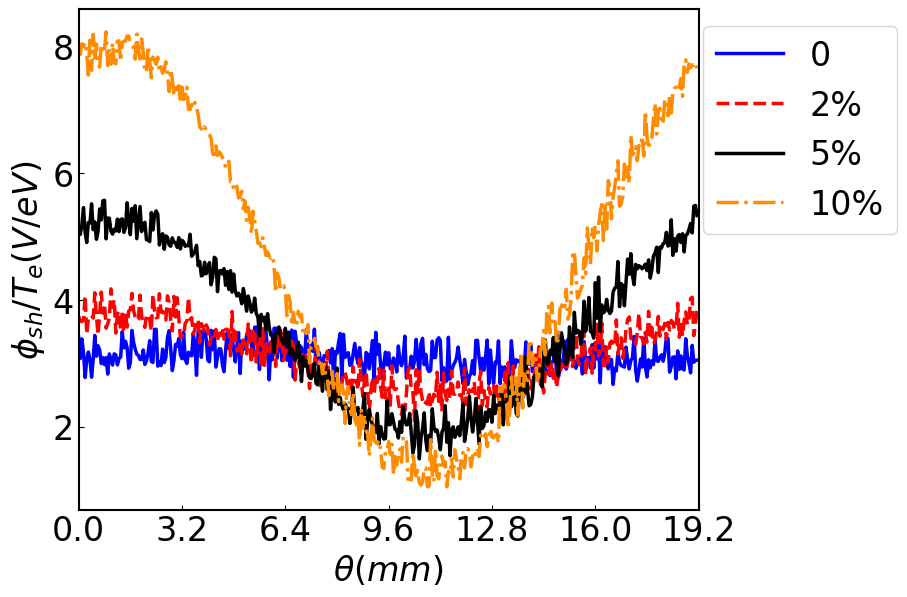}
	\caption{Azimuthal distributions of the ratio of the sheath potential drop to the mean electron temperature at 0.015 $\mu$s,  the sheath edge is defined as the point where $(n_i - n_e)/n_e$ = 0.019\cite{Beving_2022}.}
	\label{fig:qcphi}
\end{figure}

The initial space charge separation perturbs both azimuthal
fluctuations and radial MTSI intensity.
Azimuthally varying electron density modulates wall-directed electron
fluxes, inducing inhomogeneous sheath potential drops along the azimuthal direction, as illustrated in Fig.~\ref{fig:qcphi}.
These variations directly influence MTSI intensity by altering the sheath's electron reflection capability.
Furthermore, the azimuthal sheath potential distribution may excite sheath-plasma instabilities, amplifying plasma-wall interaction dynamics.

As instabilities evolve, $E_\theta$ drives ions to adopt spatial distributions akin to electrons, thereby suppressing large-scale
charge separation. This is evidenced by the gradual attenuation of the
$\lambda$-long mode amplitude, shown in  Fig.~\ref{fig:liner}.
Concurrently, ECDI development dissipates input energy via high-frequency electron trapping, suppressing $\lambda$-long mode and MTSI amplitudes.
The observed $\lambda$-long mode exhibits strong coupling to electron-ion convective transport, strongly suggesting its identification as a 
fluid-dynamic instability characteristics akin to MTSI.

\subsubsection{Effects on the nonlinear saturation phase}
\                                                                      
\newline \indent 

\begin{figure}[ht]
\centering
    \begin{subfigure}[b]{0.35\textwidth}
    \captionsetup{justification=raggedright, singlelinecheck=false}
     \includegraphics[scale=0.34]{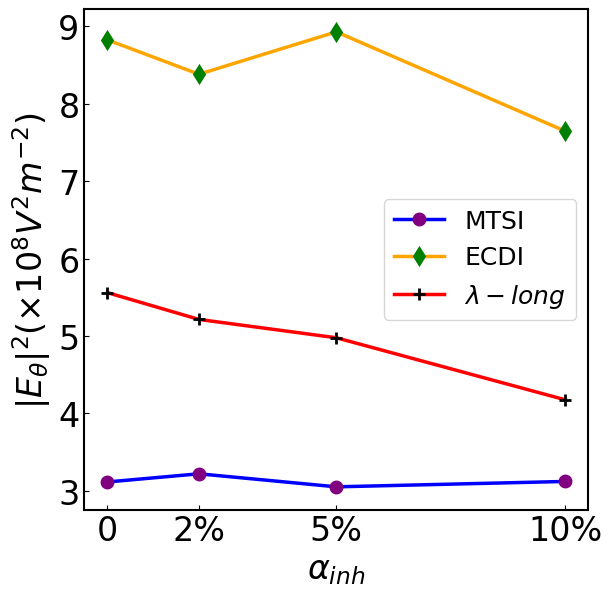}
     \caption{}
    \label{Amp}
    \end{subfigure}
    \hspace{38pt}
   \begin{subfigure}[b]{0.35\textwidth}
    \captionsetup{justification=raggedright, singlelinecheck=false}
    \includegraphics[scale=0.35]{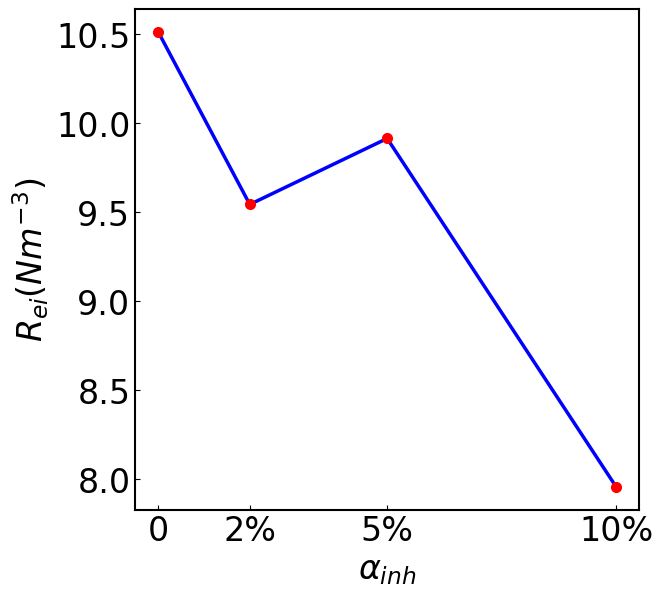}
    \caption{}
    \label{Rei}
    \end{subfigure}
\caption{ (a) Time-averaged amplitudes of each instability mode in the saturation phase within 15-30 $\mu$s. (b) The emporally and spatially averaged electron-ion relative friction within 15-30 $\mu$s obtained by Eq.~\ref{eq:Rei}.}
\label{fig:Amp_saturation}
\end{figure}

As shown in Fig.~\ref{Amp}, increased magnetic field inhomogeneity demonstrates negligible impact on MTSI.
For the $\lambda$-long mode, its intensity progressively diminishes 
with elevated magnetic field inhomogeneity, exhibiting a 24.9\% amplitude reduction under the 10\% inhomogeneity.
The ECDI amplitude remains relatively stable at 2\% and 5\% 
inhomogeneity levels but decreases by 13.4\% at 10\% inhomogeneity.
According to Ref.\cite{Lafleurcomparison}, the azimuthal electron-ion relative friction
could be obtained from the simulation results by the following equation
\begin{equation}\label{eq:Rei}
R_{ei} =q\left\langle\delta n_{e} \delta E_{\theta}\right\rangle 
\end{equation}

The relative electron-ion friction under multi-instability coupling (Fig.~\ref{Rei}) demonstrates a trend consistent with ECDI
evolution(Fig.~\ref{Amp}).
Simulation results demonstrate a 26.2\% reduction in electron-ion relative friction when magnetic field inhomogeneity reaches 10\%, indicating large friction suppression.

\begin{figure}[htb]
	\centering
	\includegraphics[width=0.45\linewidth]{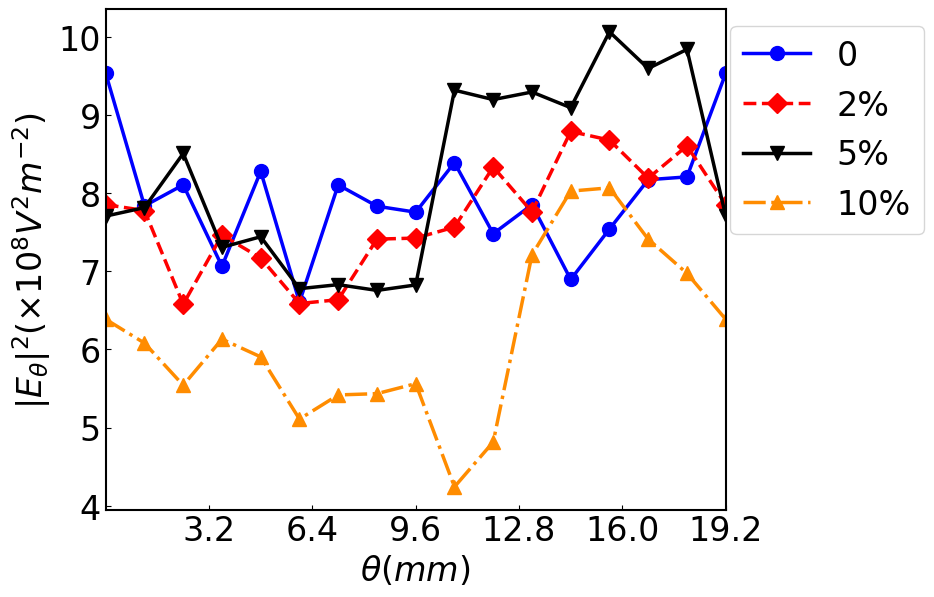}
	\caption{Azimuthal distributions of the ECDI amplitude within 15-30 $\mu$s.}
	\label{fig:ECDI_dis}
\end{figure}

The azimuthally inhomogeneous magnetic field induces inhomogeneous
electron drift velocity (as shown in Fig.~\ref{the_the}), which
exerts dual effects on ECDI intensity.
(1) The electron drift velocity directly modulates ECDI intensity.
As illustrated in Fig.~\ref{fig:ECDI_dis}, the overall azimuthal 
distribution of ECDI intensity spatially correlates with the electron
velocity profile.
As the primary energy carrier of ECDI, electrons exhibit enhanced the ECDI amplitude
in negative gradient regions, while suppressing it in positive
gradient regions due to gradient-induced effective forces
discussed in Sec.~\ref{gradient_drift}. 
Overall, $v_{de} = E/B$, $B<1$ T, increased magnetic field inhomogeneity elevates the mean azimuthal electron drift velocity, amplifying the ECDI amplitude\cite{10.1063/1.4817743}.
(2) The ECDI wavenumber couples with magnetic field strength ($k_{ECDI} = \omega_{ce}/v_{de} \propto B^2$).
The ECDI wavenumber
becomes dispersed
with enhanced field inhomogeneity, broadening its spectral width and
intensifiying nonlinear effects.
Concurrent electron thermalization (as shown in Fig.~\ref{fig:Te}) further 
amplifies spectral broadening of ECDI, leading to amplitude reduction of the ECDI discrete peak.
These dual mechanisms yield the non-monotonic in ECDI intensity with the magnetic field inhomogeneity.
\begin{figure}[htb]
	\centering
	\includegraphics[width=0.6\linewidth]{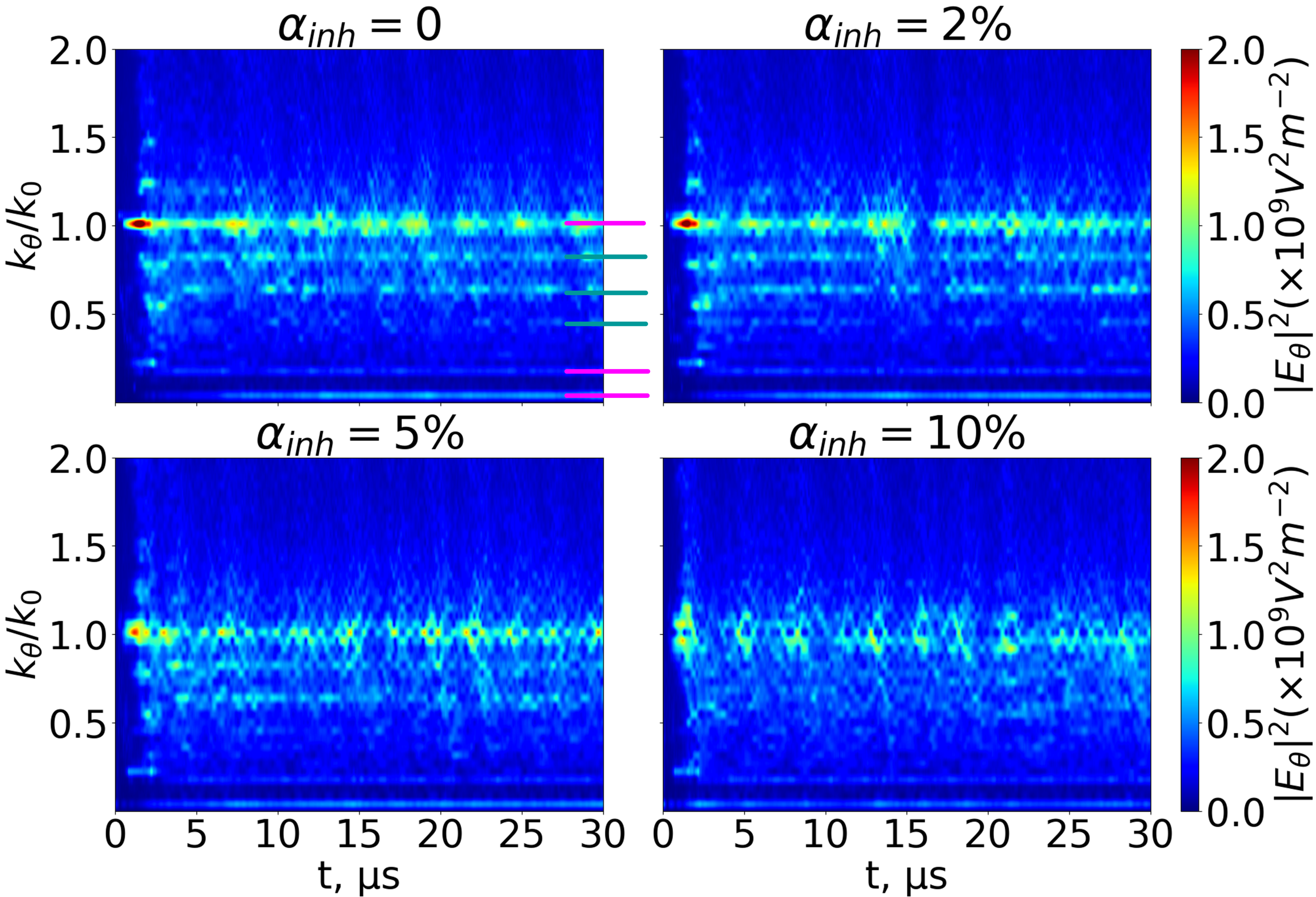}
	\caption{Temporal evolution of instability amplitudes in normalized wavenumber space. The red short lines denote spectral peaks corresponding to ECDI, MTSI, and $\lambda$-long mode in descending order of the wavenumber. Green short lines denote intermediate beat frequencies arising from nonlinear coupling.}
	\label{fig:kt_t}
\end{figure}

As shown in Fig.~\ref{fig:kt_t}, nonlinear coupling between ECDI, MTSI, and $\lambda$-long mode generates beat frequency responses at middle
wavenumbers, redistributing energy toward intermediate modes.
This resonance reduces discrete spectral peaks of individual instabilities.
With increasing magnetic field inhomogeneity, the spectral broadening
of ECDI expands the wavenumber range of beat frequency responses, 
generating more chaotic spectral features, implying the formation of a more turbulent plasma turbulence state.
This spectral energy dispersion across extended wavenumber ranges reduces the $\lambda$-long mode amplitude.
The MTSI wavenumber occupies an intermediate position between ECDI and
$\lambda$-long mode, maintaining relatively stable intensity during energy transfer due to its low amplitude characteristics.

\section{Conclusions\label{sec:conclusion}}

This study investigates the effects of azimuthally inhomogeneous magnetic field configurations on high-frequency azimuthal instabilities and electron mobility within Hall thrusters through 2D radial-azimuthal simulations.
The conclusions are as follows:
\begin{itemize}

  \item Under the fixed magnetic topology in this paper, axial electron mobility remains nearly constant when magnetic field inhomogeneity is below 5\%, but shows abrupt reduction (-15.7\%) at 10\% inhomogeneity level.
  
  \item Magnetic field and density gradient drifts cause the axial electron velocity's azimuthal distribution to exhibit spatial oscillations characterized by azimuthal length scales and induce azimuthally asymmetric energy distributions of electrons. However the effect of gradient drifts on electron mobility is almost negligible.

  \item In the initial phase, electrons rapidly respond to the inhomogeneous magnetic field, while ions are almost unaffected, leading to a violent charge separation and forming a strong azimuthal electric field.
  As the magnetic field's azimuthal inhomogeneity intensifies, amplitude enhancement occurs in both $\lambda$ - long mode and MTSI, and the linear characteristics of ECDI gradually disappear.
  
  \item In the nonlinear saturation stage, as the magnetic field azimuthal inhomogeneity increases, the electron azimuthal drift velocity distribution range becomes broader.
 This results in enhanced spectral broadening of the ECDI, increased nonlinear effect, and ultimately, the energy of each unstable mode is dispersed to a wider wavenumber range. 
 Consequently, the plasma azimuthal wavenumber spatial spectrum becomes more turbulent.
 This effect is more pronounced when the magnetic field inhomogeneity reaches 10\%, leading to a large decrease in the intensity of ECDI and $\lambda$ - long mode, and thus to a decrease in the electron mobility.

\end{itemize}

Contrary to previous findings by Bak et al\cite{10.1063/5.0067310}, our study demonstrates that azimuthal magnetic field inhomogeneity might not primarily influence electron axial mobility through high-frequency azimuthal instabilities.
The other two possible mechanisms are:
(1) As demonstrated in Sec.~\ref{sec:plasma dis}, the azimuthal magnetic field inhomogeneity induces corresponding density redistribution along the azimuthal direction, potentially creating localized variations in ionization characteristics that could enhance electron axial transport.
(2) The azimuthal magnetic field inhomogeneity might excite low-frequency instabilities in either azimuthal or axial directions.

In conclusion, the model employed in this paper is only applicable to capture the coupling evolution characteristics of high-frequency azimuthal instabilities, such as ECDI and MTSI.
Further in-depth studies on the related influence mechanisms are required in future research. 
Furthermore, this study found that the azimuthal inhomogeneous magnetic field may effect the sheath instability and the electron near-wall conductivity, which provides a direction for the subsequent research.

\section*{Acknowledgment}
The authors acknowledge the support from National Natural Science Foundation of China (Grant No.5247120164). 
This research used the open-source particle-in-cell code WarpX \href{https://github.com/ECP-WarpX/WarpX}{https://github.com/ECP-WarpX/WarpX}, primarily funded by the US DOE Exascale Computing Project. Primary WarpX contributors are with LBNL, LLNL, CEA-LIDYLSLAC, DESY, CERN, and TAE Technologies. We acknowledge all WarpX contributors.

\section*{Reference}

\bibliographystyle{elsarticle-num}
\bibliography{reference}

\end{document}